\documentclass[aps,prb,superscriptaddress,twocolumn]{revtex4-1}
\usepackage{graphicx}
\usepackage{amsmath}
\usepackage[english]{babel}
\usepackage{array}
\usepackage{amssymb, bm}
\usepackage{multirow}

\begin{document}
\title{Interface Contributions to the Spin-Orbit Interaction Parameters
of Electrons at the (001) GaAs/AlGaAs Interface}

\author{Zh.A. Devizorova}
\email{DevizorovaZhanna@gmail.com}
\affiliation
{Moscow Institute of Physics and Technology, 141700 Dolgoprudnyi, Moscow District, Russia}
\affiliation
{Kotelnikov Institute of Radio Engineering and Electronics, Russian Academy of Sciences, 125009 Moscow, Russia}

\author{A.V. Shchepetilnikov}
\affiliation
{Institute of Solid State Physics, Russian Academy of Sciences, 142432 Chernogolovka, Moscow District, Russia}
\affiliation
{Moscow Institute of Physics and Technology, 141700 Dolgoprudnyi, Moscow District, Russia}

\author{Yu.A. Nefyodov}
\affiliation
{Institute of Solid State Physics, Russian Academy of Sciences, 142432 Chernogolovka, Moscow District, Russia}

\author{V.A.Volkov}
\email{Volkov.V.A@gmail.com}
\affiliation
{Kotelnikov Institute of Radio Engineering and Electronics, Russian Academy of Sciences, 125009 Moscow, Russia}
\affiliation
{Moscow Institute of Physics and Technology, 141700 Dolgoprudnyi, Moscow District, Russia}

\author{I.V. Kukushkin}
\affiliation
{Institute of Solid State Physics, Russian Academy of Sciences, 142432 Chernogolovka, Moscow District, Russia}

\begin{abstract}
One-body mechanisms of spin splitting of the energy spectrum of 2D electrons in a one-side doped (001) GaAs/Al$_x$Ga$_{1-x}$As quantum well have been studied theoretically and experimentally. The interfacial spin splitting has been shown to compensate (enhance) considerably the contribution of the bulk Dresselhaus
(Bychkov-Rashba) mechanism. The theoretical approach is based on the solution of the  effective mass equation in a quasitriangular well supplemented by a new boundary condition at a high and atomically sharp heterobarrier. The model takes into account the spin-orbit interaction of electrons with both bulk and interfacial crystal potential having C$_{2v}$ symmetry, as well as the lack of inversion symmetry and nonparabolicity of the conduction band in GaAs. The effective 2D spin Hamiltonian including both bulk and interfacial contributions to the Dresselhaus ($\alpha_{BIA}$) and Rashba ($\alpha_{SIA}$) constants has been derived. The analytical relation between these constants and the components of the anisotropic nonlinear $g$-factor tensor in an oblique quantizing magnetic field has been found. The experimental approach is based, on one hand, on the detection of electron spin resonance in the microwave range and, on the other hand, on photoluminescence measurements of the nonparabolicity parameter. The interfacial contributions to $\alpha_{BIA}$ and $\alpha_{SIA}$ have been found from comparison with the theory.
\end{abstract}
\maketitle

\section{Introduction.}
There exist two contributions linear in the two-dimensional (2D) momentum to the Hamiltonian of conduction band electrons in (001) GaAs/Al$_x$Ga$_{1-x}$As
heterojunctions and quantum wells with a 2D electron gas. The first one, the Dresselhaus contribution [1], is caused by the lack of inversion symmetry in the bulk material (BIA) and characterized by the constant $\alpha_{BIA}$.
The second one, the Bychkov-Rashba contribution [2], appears in asymmetric structures (SIA) and is characterized by the constant $\alpha_{SIA}$. These constants are proportional to the bulk parameters $\gamma_c$ and $a_{SO}$, respectively. 

Large number of theoretical and experimental works devoted  both to calculation and measurement of the indicated parameters in the quantum wells of the above type by various techniques have been published so far (see reviews [3.5]). Many publications mentioned a large spread in the experimental values and their disagreement with the theoretical results. The parameter $|\gamma_c|$ varies from 26 [6] and 28 [7] to 11 [8] eV \AA$^3$ (see also the consolidated table in [9]). Paradoxically, the range of spread increases with the use of more and more sophisticated experimental techniques. In particular, the  values of $|\gamma_c|$ found in recent works from comparison with existing theories are (in the same units) 6 [10], 5 [11, 12], 11 [13, 14], 12 [15], and even 3 [16]. There exist much fewer experimental data on $a_{SO}$ but their spread is also impressing: from 4 [17] to 7 [11] and even 25 [18] \AA $^2 / \hbar$. At the same time, the spread in the data on ƒÁc in bulk GaAs samples is pretty small. For example, the result $|\gamma_c| = 24.5$ eV \AA$^3$ of one of the early works [19] agrees with the kp-theory (see also [20]) and was not significantly revised for 30 years. The sign of $\gamma_c$ deserves separate discussion.

This state of the research cannot be regarded as satisfactory. A number of works [4, 5, 9, 11] indicated the possibility  of the interfacial spin-orbit
interaction as a possible reason. Such interaction can be, in particular, associated with the inversion asymmetry of the interface. However, the attempts to constract the theory of the interface spin-orbit interaction consistent with the experiment encounter difficulties associated with both lack of information on the structure of an atomically sharp heterointerface at the atomic scale and inapplicability of the envelope functions approximation to the heterointerface (for details and erences, see [21-23]). In this work, we attempt to construct the phenomenological theory of the interfacial spin-orbit interaction of the general form which describes the experiment quantitatively.

In Section 2, we develop the theory of the spin-orbit splitting of the spectrum of 2D electrons at the heterojunction in the absence of a magnetic field. We
use the generalization of the single-band 3D effective mass method for the case of a high heterobarrier. The atomically sharp interface is described by a new
boundary condition for the envelope functions, which includes the contributions of the interfacial spin-orbit interaction of the general form. The deviation of this boundary condition from the zero one is considered as a small parameter of the theory. We derive the renormalized 3D Hamiltonian for the envelope functions satisfying this boundary condition in the lowest order in the above parameter and the magnitude of the spin-orbit interaction. Subsequent averaging of this Hamiltonian over the motion normal the interface leads to the sought 2D Hamiltonian containing the renormalized parameters $\alpha_{BIA}$ and $\alpha_{SIA}$. A similar approach was implemented in [23] for the simple special case of the interfacial spin-orbit interaction. However, the agreement with the experiment was not achieved therein. In Section 3, we study the spin splitting in an oblique magnetic field and derive the expressions for the components of the $g$-factor tensor including the interfacial
renormalizations.

Sections 4 and 5 are devoted to the experimental verification of the elaborated theory in relatively wide one-side doped quantum wells. The motion of electrons in such quantum wells is limited by the (001) GaAs/Al$_x$Ga$_{1-x}$As interface from one side and by the internal electric field from the other side. The components of the $g$-factor tensor and their derivatives with respect to the quantizing component of the magnetic field were measured by ESR (Section 4). Photoluminescence from the quantum wells of different widths was studied, therefrom the nonparabolicity parameter of the conduction band of GaAs was determined (Section 5). In Section 6, we compare the theory and the experiment and extract the spin parameters of the problem. The results are discussed in Section 7.

\section {THEORY OF SPIN SPLITTING OF 2D ELECTRONS SPECTRUM IN ZERO MAGNETIC FIELD.} 

We will consider a one-side doped GaAs quantum well as a quasi-triangular potential well $V(z)$ with electrons in the well pressed to the (001) GaAs/Al$_x$Ga$_{1-x}$As heterointerface by the internal electric field ${\bf F} =(0,0,F)$. Let the external normal to the interface be directed along the $z$ axis. The $x$, $y$, and $z$ axes are oriented along the crystallographic directions $[100]$, $[010]$, and $[001]$, respectively. We will assume the band discontinuity at the interface ($z = 0$) to be high and atomically sharp and the heterointerface to be impenetrable. The dynamics of an electron outside the interface ($z < 0$) will be described by the single-band effective mass equation for the two component spinor $\phi$ composed of two envelope functions. At $z = 0$, this spinor must obey the "single-side" phenomenological boundary condition containing information on the microscopic structure of the interface. This formulation of the problem is not new (see, e.g., [24–27]). What is new is the boundary condition for the envelope functions obtained by generalizing the approach [23] and taking into account both the interface and bulk spin-orbit interactions.

The 3D Hamiltonian of the conduction band in the effective mass approximation includes the contributions $H_{BIA}$ and $H_{SIA}$ and describing the spin splitting owing to the lack of inversion symmetry of the crystal and asymmetry of the well:
\begin{equation}
\label{ham}
\hat H=\frac{{\hat p}^2}{2m^*}+V(z)+\hat H_{BIA}+\hat H_{SIA},
\end{equation}
\begin{equation}
\hat H_{BIA}=\frac{\gamma_c}{\hbar^3 } \biggl[\sigma_x p_x (p_y^2-\hat p_z^2)+\sigma_y p_y (\hat p_z^2-p_x^2)+\sigma_z \hat p_z (p_x^2-p_y^2)\biggr],
\end{equation}
\begin{equation}
\hat H_{SIA}=a_{SO}(\sigma_x p_y - \sigma_y p_x)\partial_z V(z),
\end{equation}
where $\sigma_x$, $\sigma_y$, $\sigma_z$ are the Pauli matrices.

The requirements of the Hermiticity of Hamiltonian (1) in the half-space and the time reversal invariance of the problem lead to the boundary condition for
the envelope functions, which includes seven (one scalar and six spin)  phenomenological parameters. However, it follows from the C$_{2v}$ symmetry that the spin part of the Hamiltonian [28] and the boundary condition is described (in the lowest order in the 2D momentum) only by the Dresselhaus and Bychkov-Rashba contributions. This allows one to reduce the number of boundary parameters to three.

Thus, the boundary condition for the spinor ö taking into account the spin–orbit interaction in the bulk and at the interface with the C$_{2v}$ symmetry, as well as the inversion asymmetry of the bulk crystal, has the form 
\begin{multline}
\label{GU}
\Biggl. \Biggl[\sigma_0 - i\frac{R \hat p_z}{\hbar} -i\frac{2m^* \gamma_c R}{\hbar^4}(\sigma_y p_y -\sigma_x p_x) \hat p_z  - i\frac{m^* \gamma_c R}{{\hbar^4}} \sigma_z (p_x^2 - p_y^2)+\\+ \frac{(\chi+\chi^{int}) R}{\hbar}(\sigma_x p_y - \sigma_y p_x)- \frac{2m^*\gamma_c^{int}}{\hbar^3}(\sigma_y p_y - \sigma_x p_x)\Biggr]\phi \Biggr|_{z=0}=0.
\end{multline}
Here, the real quantity $R$ depends on the microscopic structure of the interface, $\sigma_0$ is the identity matrix, $\gamma_c$ and $\chi$ are the bulk parameters (for GaAs, $|\gamma_c|$= 24,4 eV $\times$ \AA$^3$, $\chi=0.082$), and the constants $\gamma_c^{int}$ and $\chi^{int}$ characterize the spin-orbit interaction at the interface. The special case of $\gamma_c^{int} =\chi^{int}= 0$ was considered in [23].

We will assume that the difference between boundary condition (4) and the zero one is small. This allows one to use perturbation theory to the lowest order in $R$ and the parameters of the spin-orbit interaction. Following [23, 25], we reduce the problem with Hamiltonian (1) and boundary conditions (4) to a new simpler problem with zero boundary conditions and the Hamiltonian renormalized owing to the interfacial contributions. Averaging the renormalized 3D Hamiltonian over the lowest subband functions, we find the effective 2D Hamiltonian, which includes the Dresselhaus and Bychkov-Rashba terms. The respective constants $\alpha_{BIA}$ and $\alpha_{SIA}$, apart from the known bulk contributions, include both the scalar contribution of the interface ($R$) and the renormalized parameters of the spin-orbit interaction of the Dresselhaus ($\tilde{\gamma_c}=\gamma_c+\gamma_c^{int}$) and Rashba ($\tilde{\chi}=\chi+\chi^{int}$) types:
\begin{equation}
\hat H_{2D}=\frac{p_x^2+p_y^2}{2m^*}+\alpha_{BIA} (\sigma_y p_y - \sigma_x p_x) + \alpha _{SIA} (\sigma_x p_y -\sigma_y p_x),
\end{equation}
\begin{equation}
\label{Dress const}
\alpha_{BIA} = \frac{\gamma_c}{\hbar ^3} (\hat p_z ^2)_{00}+\frac{2 m^* \tilde{\gamma_c}} {\hbar ^3} e F R,
\end{equation}
\begin{equation}
\label{Rashba const}
\alpha_{SIA} = -eF \left( a_{SO}+\frac{\tilde{\chi} R^2}{\hbar} \right),
\end{equation}
where $F=-(\partial_z V/e)_{00}$ is the average electric field in the heterojunction and $-e$ is the elementary charge. In contrast to [23], Eqs. (6) and (7) include the contributions from the interface spin-orbit interaction to the Dresselhaus ($\gamma_c^{int}$) and Rashba ($\chi^{int}$) constants.

\section{THEORY OF SPIN SPLITTING OF THE 2D ELECTRON SPECTRUM IN AN OBLIQUE MAGNETIC FIELD}
Following [23], we write Hamiltonian (1) and boundary conditions (4) in an oblique magnetic field. In contrast to [23], we take into account the effect of
nonparabolicity of the conduction band of GaAs on the Zeeman splitting. The emerging contribution to the spin part of Hamiltonian (1) has the form [29]
\begin{equation}
\label{np in mf}
\hat H_{np}= \frac{1}{2} \mu_B \frac{h_1 (\hat p-eA/c)^2}{\hbar^2}(\bm {\sigma B}),
\end{equation}
where $\mu_B$ is the Bohr magneton and  $h_1$ is the nonparabolicity parameter governed by the band structure of the bulk semiconductor. The corresponding contribution
to boundary conditions (4) is $-i\mu_B m^*h_1R \hat p_z (\bm{\sigma} {\bf B})  /  \hbar^3$.

After calculations similar to those performed in [23], we find the tensor $g_{\alpha \beta}(B)=g_{\alpha \beta}(0)+d_{\alpha \beta}|B_z|$, which is anisotropic in the $\alpha, \beta=x,y$ plane. Its components are given by the expressions
\begin{equation}	
\label{gzz in zero mf with np}
g_{zz}(0)=g^*+\frac{2m^*h_1}{\hbar^2}\left(\frac{(p_z^2)_{00}}{2m^*}+eFR\right),
\end{equation}
\begin{equation}
\label{dzz with np}
d_{zz}=\frac{2eh_1}{c\hbar }\left(N+\frac{1}{2}\right),
\end{equation}
\begin{multline}
\label{gxx in zero mf with np}
g_{xx}(0)=g_{yy}(0)=g^*+\frac{4\tilde{\chi} R^2m_0}{\hbar^2}\frac{(p_z^2)_{00}}{m^*}+\\+\frac{2m^*h_1}{\hbar^2}\left(\frac{(p_z^2)_{00}}{2m^*}+eFR\right),
\end{multline}
\begin{equation}
\label{dxx with np}
d_{xx}=d_{yy}=\frac{2eh_1}{c\hbar }\left(N+\frac{1}{2}\right)-\frac{4 \tilde{\chi} R^2 m_0e}{m^*c\hbar}\left(N+\frac{1}{2}\right),
\end{equation}
\begin{multline}
\label{gxy in zero mf with np}
g_{xy}(0)=g_{yx}(0)=\frac{4m_0\gamma_c}{\hbar ^4}  [ (p^2_z)_{00} z_{00}-(p^2_z z)_{00}]-\\-\frac{8m_0 m^* R\tilde{\gamma_c}}{\hbar^4}\frac{(p_z^2)_{00}}{m^*},
\end{multline}
\begin{equation}
\label{dxy with np}
d_{xy}=d_{yx}=\frac{8m_0e\tilde{\gamma_c}R}{c\hbar^3}\left(N+\frac{1}{2}\right),
\end{equation}
where $g^*=-0.44$ and $N$ is the number of the Landau level.

The terms in Eqs. (9)--(14) proportional to $R$, $\gamma_c^{int}$, $\chi^{int}$, and their combinations are the sought interface contributions. If they vanish, results (9)–(14) are reduced to the known ones [29–31]. 

The constants $\alpha_{BIA}$ and $\alpha_{SIA}$  can be expressed in terms of the components of the tensor $g_{\alpha \beta}$:
\begin{multline}
\label{alpha BIA over gij}
\alpha_{BIA} = \frac{\gamma_c}{\hbar ^3} (\hat p_z ^2)_{00}-\\-e F\frac{m^*\hbar}{4m_0(p_z^2)_{00}}\left(g_{xy}-\frac{4m_0\gamma_c}{\hbar^4}( (p^2_z)_{00} z_{00}-(p^2_z z)_{00})\right),
\end{multline}
\begin{equation}
\label{alpha SIA over gij}
\alpha_{SIA} = -eF \left( a_{SO}+\frac{m^*\hbar}{4m_0(p_z^2)_{00}}(g_{xx}-g_{zz})\right).
\end{equation}

\section{EXPERIMENT: ELECTRON SPIN RESONANCE}
As follows from the above one-body theory, the correct determination of all parameters describing the spin splitting in the two-dimensional electron system
requires sufficiently accurate measurement of all components of the one-body nonlinear $g$-factor tensor. The most appropriate experimental technique is the electron spin (paramagnetic) resonance (ESR). The one-body $g$-factor was determined by this method (for details, see [32, 33]).

\begin{figure}[t]
\centerline{\includegraphics[width=0.95\columnwidth,clip]{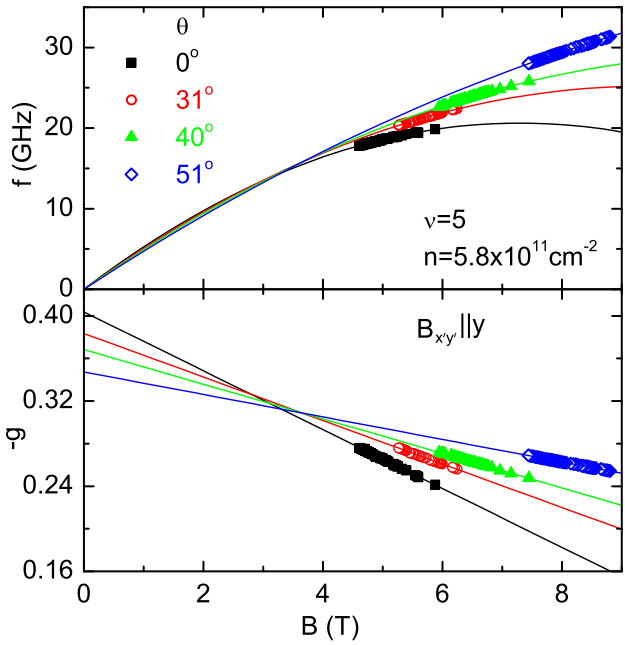}}
\caption{Magnetic field dependence of the  ESR frequency (upper panel) and  $g$- factor (lower panel) for four different values of the angle $\theta$ between the magnetic field and the normal to the plane of the two-dimensional electron system. The in-plane component $B_{x'y'}$ of the magnetic field lies along the $Oy'$ axis. The $g^*$ values are found as extrapolations of the experimental data to zero magnetic field (solid lines). } \label{f_g_B}
\end{figure}

\begin{figure}[t]
\centerline{\includegraphics[width=0.95\columnwidth,clip]{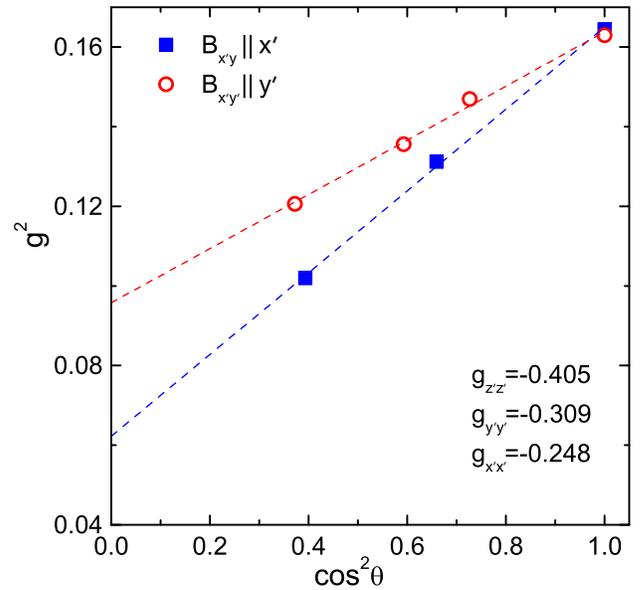}}
\caption{Square of the effective $g$ factor versus $cos^2\theta$ for two orientations of the in-plane component of the magnetic field: $B_{x'y'} || x'$ (squares) and $B_{x'y'} || y'$ (circles). The quantities $g_{x'x'}$ and $g_{y'y'}$ were calculated with the use of Eq. (7) from the intercepts of the dashed lines with the vertical axis.}
\label{cos2}
\end{figure}

Charged AlGaAs/GaAs heterojunctions or wide AlGaAs/GaAs quantum wells with a high built-in electric field, i.e., with a high electron density, are most appropriate for the experimental verification of the calculations.

We chose a one-side $\delta$-silicon doped 25 nm (001)AlGaAs/GaAs quantum well for the experiments. The structure was grown by molecular beam epitaxy. The electron density and mobility in the structure at the temperature $T = 1.3$ K were $n=5.8\times 10^{11}$ cm$^2$ and $\mu=1.3\times 10^6$ V/(cm$^2$ $\times$ s). The sample was a mesa structure with the shape of a standard Hall bar with source, drain, and potentiometric contacts.

We implemented an indirect detection of ESR [32] based on the ultimate sensitivity of the resistance $R_{xx}$ of the 2D electron system in the quantum Hall effect regime to the microwave absorption. Below, we briefly describe the measurement technique (for details, see [33.35]). The measurement of the variation $\delta R_{xx}$ induced by the absorption of the microwave radiation
was performed by double lock-in detection. An alternating current with the frequency $f \sim 1$ kHz and the  amplitude $I_{RMS} \sim 1$ $\mu$A passed from the source to the drain. The sample was irradiated by the 1~mW micro wave radiation amplitude modulated at the frequency $f_{mod} \sim 30$ Hz. The first lock-in detector was tuned to the frequency of the ac current and measured the signal proportional to the resistance $R_{xx}$. The second lock-in detector took the signal from the output of the first detector and was tuned to the modulation frequency of the microwave radiation; i.e., it measured the variation $\delta R_{xx}$ of the resistance induced by the microwave absorption. We kept the frequency of the incident radiation constant and swept the magnetic field ${\bf B}$, thus measuring the dependences $R_{xx}(B)$ and $\delta R_{xx}(B)$. The dependence $\delta R_{xx}(B)$ exhibited a peak corresponding to the electron spin resonance. The experiments were carried out at the temperature $T=1.3$ K in magnetic fields up to $10$ T. The ESR signal was observed near the filling factors $\nu=3,5,7$. The measurements were largely performed near $\nu=5$.

The sample was mounted on a rotating substrate. Thus, it was possible to vary {\it in situ} the angle $\theta$ between the normal to the $[001]$ plane and the direction of the magnetic field and the angle $\phi$ between the $[110]$ direction and the in-plane component of the magnetic field. The angles were controlled by the three-dimensional Hall probe firmly attached to the sample holder.

We measured the dependence of the ESR frequency $f$ on the magnetic field ${\bf B}$ at the fixed angle $\theta$ and $\phi$. From these data, we computed the dependences $g(B)$. The dependences $f(B)$ and $g(B)$ for the angles $\theta=$0, 31, 40 and $51^\circ$ and the fixed angle $\phi=90^\circ$ are shown in Fig. 1. The dependences $g(B)$ are linear near an odd filling factor [36]. The linear extrapolation of $g(B)$ to the zero magnetic field allows one to find the one-body $g$-factor $g_0(\theta,\phi)$ near the bottom of the quantum confinement subband. 

Let the axes Ox', Oy', and Oz be oriented along the $[110]$, $[1\overline{1}0]$ and $[001]$ crystallographic directions, respectively. Then, the following equation holds:
\begin{equation}
g_0^2(\theta,\phi)={[g_{x'x'}^2\cos^2\phi+g_{y'y'}^2\sin^2\phi]\sin^2{\theta}
+g_{zz}^2\cos^2\theta}\label{g*}.
\end{equation}
Thus, if one fixes the angle  $\phi=0^\circ$ or $\phi=90^\circ$, the linear extrapolation of the experimental dependence $g^2 (cos^2\theta)$ allows one to calculate $g_{x'x'}$ or $g_{y'y'}$, respectively. The measured dependences $g^2 (cos^2\theta)$ for $\phi=0^\circ$ (squares) and $\phi=90^\circ$ (circles) are shown in Fig. 2. As is seen, they are linear, in full agreement with Eq. (17). The intercepts of these dependences with the vertical axis yield $g_{x'x'}=-0.248 \pm 0.005$ and $g_{y'y'}=-0.309 \pm 0.005$. The  value $g_{zz} =-0.405 \pm 0.001$ was determined directly  from the measurements in the transverse field. Following [34, 35], we found the corrections to the g factor linear in the magnetic field (near $\nu=5$): $d_{zz}=0.0278\pm0.0005$~T$^{-1}$,
$d_{y'y'}=0.010\pm0.001$~T$^{-1}$, $d_{x'x'}=-0.001\pm0.001$~T$^{-1}$.

\section{EXPERIMENT: PHOTOLUMINESCENCE MEASUREMENTS OF THE CONDUCTION BAND
NONPARABOLICITY PARAMETER IN GaAs}
According to Eq. (9), the nonparabolicity parameter $h_1$ describes the dependence of $g_{zz}$ on the quantum confinement energy $\Delta E$ in the quantum wells. For its  experimental determination, we measured $g_{zz}$ in a series of charged GaAs/AlGaAs quantum wells characterized by different values of $\Delta E$ (different well widths and potential barrier heights). The width of the quantum wells varied from $7$~nm äî $25$~nm and the concentration $x$ of Al in the barrier layers ranged from $15\%$ to $32\%$. All studied quantum wells were grown along the $[001]$ direction by molecular beam epitaxy.

\begin{figure}[t]
\centerline{\includegraphics[width=0.95\columnwidth,clip]{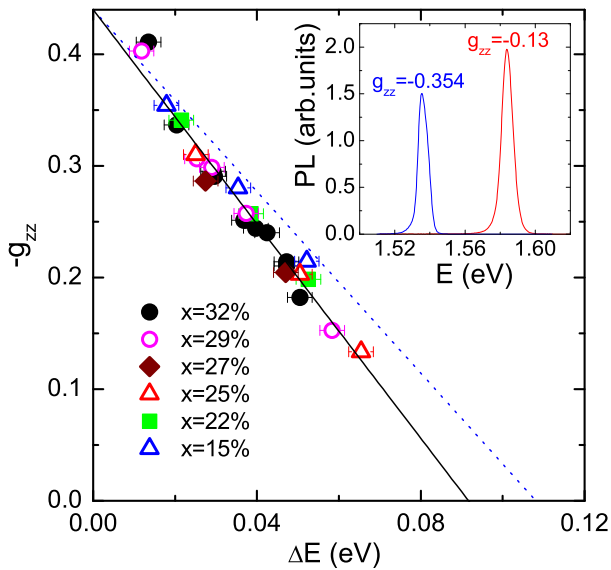}}
\caption{Component $g_{zz}$ versus the quantum confinement energy in the series of GaAs/Al$_x$Ga$_{1–x}$As quantum wells with different widths and concentrations $x$ of aluminum. The solid line is the fit of the experimental
data to Eq. (9) in the rectangular well approximation at $2mh_1/\hbar^2=4.8\pm0.6$~eV$^{-1}$ . The dotted line corresponds to $2mh_1/\hbar^2=4.1$~eV$^{-1}$ calculated in the multiband Kane model. The inset exemplifies the photoluminescence
spectrum.} \label{PL}
\end{figure}

The quantum confinement energy $\Delta E$ in such wells was determined from the photoluminescence spectra. Photoexcitation was performed using laser light with a wavelength of $532$ nm. The photoluminescence spectrum is exemplified in the inset in Fig. 3. The green light incident on the sample considerably decreased
the electron density in the well, thus red shifting the photoluminescence peak. This effect is significant in wide quantum wells. However, the shape of the narrow quantum wells can be regarded as rectangular with a reasonable accuracy. A change in the electron density in such wells hardly affects $\Delta E$. Consequently, this effect can be neglected. The hole mass in GaAs/AlGaAs is much larger than the electron mass. Therefore, the contribution of the quantum confinement of holes to the position of the photoluminescence peak can be disregarded.

The results of measurements are shown in Fig. 3. As is seen, two points with the lowest ÄE values corresponding to the quantum wells with the width larger
than $20$ nm fall out of the general linear dependence $g_{zz}(\Delta E)$. The other experimental points were obtained for the narrower wells with the widths not exceeding $14$ nm. The respective points fit well the general linear dependence predicted by Eq. (9) in the rectangular well approximation. The linear fit of the experimental dependence $g_{zz}(\Delta E)$ to Eq. (16) allows one to find the nonparabolicity parameter: $2mh_1/\hbar^2=4.8\pm0.6$~eV$^{-1}$., which is close to $2mh_1/\hbar^2=4.1$~eV$^{-1}$ calculated theoretically in the Kane model.

\section{COMPARISON OF THE THEORY AND EXPERIMENT}
We calculated the constant $\gamma_c$ in the extended (14-band) Kane model in the fourth-order kp perturbation theory to be $|\gamma_c| = 24.4$ eV $\times$ \AA $^3$. The absolute value of $\gamma_c$ agrees perfectly well with the bulk data [19]. However, there is disagreement in the literature regarding the
sign of $\gamma_c$. Many authors found $\gamma_c>0$ [3, 5, 37], whereas some other found $\gamma_c<0$ [12, 13].

Let us compare the theory with the data of Section 5 for both possible signs of $\gamma_c$. First, we transform the components of the tensors from the principal axes to the cubic ones: $F=0.4 \times 10^5$ V/cm, $g_{xx}=-0.2785$, $g_{zz}=-0.405$, $g_{xy}=0.0305$, $d_{xx}/(N+1/2)=0.0018$ T$^{-1}$, $d_{zz}/(N+1/2)=0.0112$ T$^{-1}$, $d_{xy}/(N+1/2)=-0.0022$ T$^{-1}$. 

We start from the largest components of the $g$-factor tensor. First, we process the dependence given by Eq. (9), wherefrom we find $R=22$ \AA. Then, knowing $h_1$, we calculate the theoretical value $d_{zz}/(N+1/2)=0.008$  T$^{-1}$ and compare it with the experimental one. Next, we find the combination $\tilde{\chi}R^2$, which enters into Eqs. (11) and (12). Now, we can find $d_{xx}/(N+1/2)=0.0016$ T$^{-1}$ with the use of Eq. (12) and $\tilde{\chi}=0.015$. At last, we process the off-diagonal component $g_{xy}(0)$ and find the combination $\tilde{\gamma_c}R$ from the comparison of Eq. (13) with the experimental value of $g_{xy}$. We estimate $d_{xy}/(N+1/2)$ according to Eq. (14) using the value R and find . In addition, we estimate the constants $\alpha_{BIA}$ and $\alpha_{SIA}$. The results are presented in Tables I and II.

\begin{table}[ht]
\begin{center}
{\footnotesize
\begin{tabular}{| c  | c | c | c | c |}
  \hline    
\multicolumn{3}{ |c| }{ $R=-22$ \AA} & \multicolumn{2}{ c|}{$2m^*h_1/\hbar^2=4.8$ 1/eV}  \\ \hline 
$\tilde{\gamma_c}=\gamma_c+\gamma_c^{int}$  &  $\gamma_c>0$ &  $\gamma_c<0$   & \multicolumn{2}{ c| }{$\tilde{\chi}=\chi+\chi^{int}$}  \\  \hline
$\gamma_c$ (eV$\times$ \AA$^3$) & 24.4 &   -24.4   & $\chi$ & 0.082 \\ 
$\tilde{\gamma_c}$ (eV$\times$ \AA$^3$) & 22.2 &  -13.2   & $\tilde{\chi}$ & 0.015\\ 
$\gamma_c^{int}$ (eV$\times$ \AA$^3$) & -2.2 &   11.2   &$\chi^{int}$ & -0.067 \\ \hline
\multicolumn{3}{ |c| }{$\alpha_{BIA}=\alpha_{BIA}^{(0)}+\alpha_{BIA}^{int}$} & \multicolumn{2}{ c| }{$\alpha_{SIA}=\alpha_{SIA}^{(0)}+\alpha_{SIA}^{int}$} \\ \hline
$\alpha_{BIA} \times \hbar$ (meV) & 3.5 &  -4.9  & $\alpha_{SIA}\times \hbar$ (meV) & -4.8\\
$\alpha_{BIA}^{(0)} \times \hbar$ (meV) & 7 &  -7    & $\alpha_{SIA}^{(0)} \times \hbar$ (meV) & -1.9\\ 
$\alpha_{BIA}^{int} \times \hbar$ (meV) & -3.5 &  2.1  & $\alpha_{SIA}^{int} \times \hbar$ (meV) & -2.9\\
\hline 
\end{tabular}}
\caption{}
\end{center}
\end{table}

\begin{table}[ht]
\begin{center}
\begin{tabular}{| c | c | c |c |}
  \hline             
  &  Exp. (T$^{-1}$)  & Theor. (T$^{-1}$) \\ \hline 
$d_{zz}/(N+1/2)$  &0.0112 & 0.0083  \\ \hline 
$d_{xx}/(N+1/2)$  &0.0018 &0.0016  \\\hline 
\multirow{2}{*}{$d_{xy}/(N+1/2)$}  &\multirow{2}{*}{ -0.0022}& -0.008 ($\gamma_c>0$) \\
     &    & 0.0047 ($\gamma_c<0$) \\
\hline
\end{tabular}
\caption{}
\end{center}
\end{table}

\section{DISCUSSION OF THE RESULTS AND CONCLUSIONS}
In this work, we have constructed the one-body theory of interface spin-orbit interaction with the use of the envelope function method. The theory describes
quantitatively the experimental data on ESR in the system of 2D electrons pressed by the internal electric field to one heterointerface in a wide quantum well. Below, we summarize the results.

The boundary condition for the envelope functions at the atomically sharp impenetrable (001) GaAs/AlGaAs interface have the form of Eq. (4). It takes into account the spin-orbit interaction both in the bulk and at the interface, the inversion asymmetry of the crystal, and the C$_{2v}$ symmetry of the interface. Next, we have used the smallness of the difference of its boundary condition from zero one. Expressions (9), (11), and (13) for the components of
the $g$-factor tensor and Eqs. (10), (12), and (14) for its derivatives with respect to $|B_z|$ taking into account spin-orbit interaction, interface spin-orbit interaction, asymmetry of the quantum well,  nonparabolicity of the conduction band and including a considerable interfacial renormalization have been derived in the lowest order in the spin-orbit interaction. The quantities $\alpha_{BIA}$ and $\alpha_{SIA}$ have been expressed in terms of the $g$-factor tensor components with the use of Eqs. (15) and (16).

The expressions for $d_{zz}$ and $d_{xx}$ known from the literature coincide owing to the neglect of the interfacial contribution. They cannot describe the experimental results [34, 35] exhibiting a considerable difference  between $d_{zz}$ and $d_{xx}$. This difference can be explained by the interfacial contribution to $d_{xx}$ (the second term in Eq. (12)). In addition, the experiments [34, 35] revealed the off-diagonal component $d_{xy}$, which is determined by the interfacial contribution solely, as is seen from Eq. (14). Thus, the constructed theory at least qualitatively describes the experiment [34, 35].

For quantitative verification of the theory, we have studied by ESR the spin splitting of the electron spectrum in a wide GaAs/AlGaAs quantum well with a
high built-in electric field, which provides the applicability of the theory. The quantitative description of the experimental data is possible only with the inclusion of all interfacial contributions to the $g$-factor and $d_{\alpha \beta}$ (9)–(14). Comparison of theoretical expressions (9), (11), and (13) for the components of the $g$-factor tensor with the experimental data has yielded the interfacial parameters, which appear to be comparable to the bulk contributions (see Table I). Both possible signs of $\gamma_c$ have been taken into account. These parameters have been used for the quantitative description of the derivatives $d_{\alpha \beta}$. The results obtained agree well with the experimental data for $d_{zz}$ and $d_{xx}$. The sign and absolute value of $d_{xy}$ depend on the choice of the sign of the bulk $\gamma_c$ (see Table II). For $\gamma_c>0$, as in [3, 5, 37], $d_{xy}$ is negative but almost 4 times larger in magnitude than the experimental value, which can be attributed to the
large error in finding the small quantity $g_{xy}$. If $\gamma_c<0$, as in [12, 13], $d_{xy}$ is positive.

 A large scatter of the literature experimental data on spin constants can be associated with incomplete inclusion of the interfacial spin-orbit interaction. Most often, some effective quantities containing information on the interface rather than the bulk parameters ($\gamma_c$ è $a_{SO}$) are determined. If, on the contrary, $\gamma_c$ is found not from $\alpha_{BIA}$ but rather from the momentum cubic Dresselhaus term, which does not include the interfacial renormalization, the experiment agrees with the kp theory [7].

In this work, we have analyzed the contribution of only one interface to the spin-orbit interaction. In sufficiently narrow wells, one has to take into account the contributions of both interfaces to the interface spin-orbit interaction. This problem deserves separate consideration.

This work was supported in part by the Russian Foundation for Basic Research. The work of Zh.A.D. was supported by the Dynasty Foundation.

\end{document}